\documentclass[twocolumn]{aastex61}

\usepackage{amsmath,color,textcomp,url,graphicx,subfigure}

\newcommand{\age}{$117 \pm 26$}
\newcommand{\agefull}{$117 \pm 26$ ($\pm 13$ statistic, $\pm 22$ systematic)}
\newcommand{\agenosys}{$117 \pm 13$}

\newcommand{\progenitormass}{$7.8 \pm 0.6$}
\newcommand{\msage}{$44 \pm 7$}
\newcommand{\combinedage}{$133_{-20}^{+15}$}

\newcommand{\kms}{\hbox{km\,s$^{-1}$}}
\newcommand{\msol}{$M_{\odot}$}
\newcommand{\mjup}{$M_{\mathrm{Jup}}$}

\newcommand{\masyr}{$\mathrm{mas}\,\mathrm{yr}^{-1}$}
\newcommand{\teff}{$T_{\rm eff}$}

\received{2018 June 1}
\revised{2018 June 12}
\accepted{2018 June 19}

\submitjournal{ApJ Letters.}

\shorttitle{A WHITE DWARF IN ABDMG}
\shortauthors{Gagn\'e et al.}

\begin{document}

\title{A YOUNG ULTRAMASSIVE WHITE DWARF IN THE AB DORADUS MOVING GROUP}
\author[0000-0002-2592-9612]{Jonathan Gagn\'e}
\affiliation{Carnegie Institution of Washington DTM, 5241 Broad Branch Road NW, Washington, DC~20015, USA}
\affiliation{NASA Sagan Fellow}
\email{jgagne@carnegiescience.edu}
\author[0000-0001-5485-4675]{Gilles Fontaine}
\affil{Institute for Research on Exoplanets, Universit\'e de Montr\'eal, D\'epartement de Physique, C.P.~6128 Succ. Centre-ville, Montr\'eal, QC H3C~3J7, Canada}
\author[0000-0002-0611-0222]{Am\'elie Simon}
\affil{Universit\'e de Montr\'eal, D\'epartement de Physique, C.P.~6128 Succ. Centre-ville, Montr\'eal, QC H3C~3J7, Canada}
\author[0000-0001-6251-0573]{Jacqueline K. Faherty}
\affiliation{Department of Astrophysics, American Museum of Natural History, Central Park West at 79th St., New York, NY 10024, USA}

\begin{abstract}

We use \emph{Gaia}~DR2 kinematic data and white dwarf evolutionary models to determine that the young and ultramassive white dwarf GD~50 is a likely member of the AB~Doradus moving group.\deleted{ We use the BANYAN~$\Sigma$ Bayesian classification algorithm to determine that GD~50 has a membership probability of 99.7\%. The discovery of a single white dwarf in AB~Doradus could have been expected from a fiducial log-normal initial mass function anchored on its AFG-type members.} \replaced{We use the}{Comparison with the} Montr\'eal white dwarf evolutionary models and the MIST main-sequence lifetimes \replaced{to estimate the}{imply a} total age of \deleted{GD~50, at }\agefull\,Myr, accounting for all possible C/O/Ne core compositions\added{ and using the Pleiad white dwarf LB~1497 as a comparison benchmark}. This is the first white dwarf cosmochronology age for a nearby young moving group, and allows us to refine the age of the AB~Doradus moving group at \combinedage\,Myr by combining it with its independent isochronal age.\deleted{ Our result strengthens the suggestion that ultramassive white dwarfs can be formed from a single massive progenitor, which in the present case we estimate had a mass of \progenitormass\,\msol.} GD~50 is the first white dwarf member of the AB~Doradus moving group and is located at only 31\,pc from the Sun, making it an important benchmark to better understand the star-formation history of the Solar neighborhood.

\end{abstract}


\section{INTRODUCTION}\label{sec:intro}

The recent publication of more than a billion precise trigonometric parallaxes and proper motions in the \emph{Gaia}~Data Release 2 (\citealp{Lindegren:2018gy,2018arXiv180409365G}; \emph{Gaia}~DR2 hereafter) is instigating a revolution in several branches of astrophysics, which includes our understanding of Galactic kinematics and the star formation history of the Solar neighborhood. This large influx of high-quality kinematic data \replaced{is allowing us to discover}{is enabling the discovery of} thousands of new low-mass members in known young associations (e.g., \citealt{2018arXiv180511715G}) where the initial mass function is predicted to peak (e.g., \citealp{2005ASSL..327...41C,2011AJ....141...98B}) as well as brand new young associations of stars that were not recognized before (e.g., \citealt{2017AJ....153..257O}, \citealt{2018arXiv180409058F} and J.~Gagn\'e, J.~K.~Faherty and E.~E.~Mamajek, submitted to ApJ).

Another branch of astrophysics that is strongly affected by \emph{Gaia}~DR2 is the study of white dwarfs. These objects are too faint (absolute $G$-band magnitudes of 10--15; \citealt{2018arXiv180409378G}) to have been efficiently surveyed by large trigonometric distance missions such as Hipparcos \citep{1997AA...323L..49P}, but \emph{Gaia}~DR2 can now detect them efficiently at distances well above 100\,pc \citep{2018arXiv180409378G}. This revolution of stellar astrophysics in the Solar neighborhood opens the door to a vastly improved efficiency in the search and characterization of nearby white dwarfs at well-calibrated ages through young associations within 150\,pc of the Sun.

In this Letter, we independently recover the ultramassive DA white dwarf GD~50 as a strong candidate member of the AB~Doradus moving group (ABDMG hereafter; \citealp{2004ApJ...613L..65Z,2004ARAA..42..685Z}) with recent age estimates that range from $\sim$\,100--125\,Myr \citep{2005ApJ...628L..69L,2013ApJ...766....6B} to 130--200\,Myr \citep{2015MNRAS.454..593B}. \cite{2006MNRAS.373L..45D} used the model-predicted distance for GD~50 to estimate its $UVW$ \replaced{space velocities and found that it was}{space velocity and found that it is} similar to the Pleiades. The authors note that GD~50 may therefore have formed with the Pleiades, or with the `local association' which is now thought to be a non-coeval stream \citep{MamajekIAU} that includes the Pleiades, the ABDMG and other unrelated stars. However, they find no clear explanation for the discrepant distance of GD~50 ($\sim$\,31\,pc versus $\sim$\,135\,pc for the Pleiades) and they suggest that it might have been ejected early after its formation. The lack of precise kinematic measurements and methodologies to distinguish stars from different young associations in this large range of distances prevented a clear conclusion on the membership of GD~50. As a consequence of this ambiguity, GD~50 was missing in all compilations of candidate members of the Pleiades association and the ABDMG (e.g., \citealp{2008hsf2.book..757T,2013ApJ...762...88M,2014AA...563A..45S,2017AJ....153...95R,2018ApJ...856...23G}). \added{GD~50 was also targeted in a direct-imaging search for planetary-mass companions due to its youth and proximity, but giant planets with masses above 4\,\mjup\ at separations larger than 6.2\,au were excluded at a 5$\sigma$ significance \cite{2015AA...579L...8X}.}

We use updated GD~50 kinematics from \emph{Gaia}~DR2 and the BANYAN~$\Sigma$ Bayesian classification algorithm \citep{2018ApJ...856...23G} to demonstrate that GD~50 is a likely member of the ABDMG (Section~\ref{sec:ident}), and has a negligible Pleiades membership probability. We show that the discovery of a single white dwarf member in the ABDMG is consistent with a log-normal initial mass function anchored on its AFG-type members. This determination is strengthened by an investigation of the total age of GD~50, which also provides us with the first white dwarf-based age determination for a young moving group of the Solar neighborhood (Section~\ref{sec:age}). A conclusion is presented in Section~\ref{sec:conclusion}.

\section{MEMBERSHIP}\label{sec:ident}

\begin{figure*}
	\centering
	\subfigure[$UV$]{\includegraphics[width=0.48\textwidth]{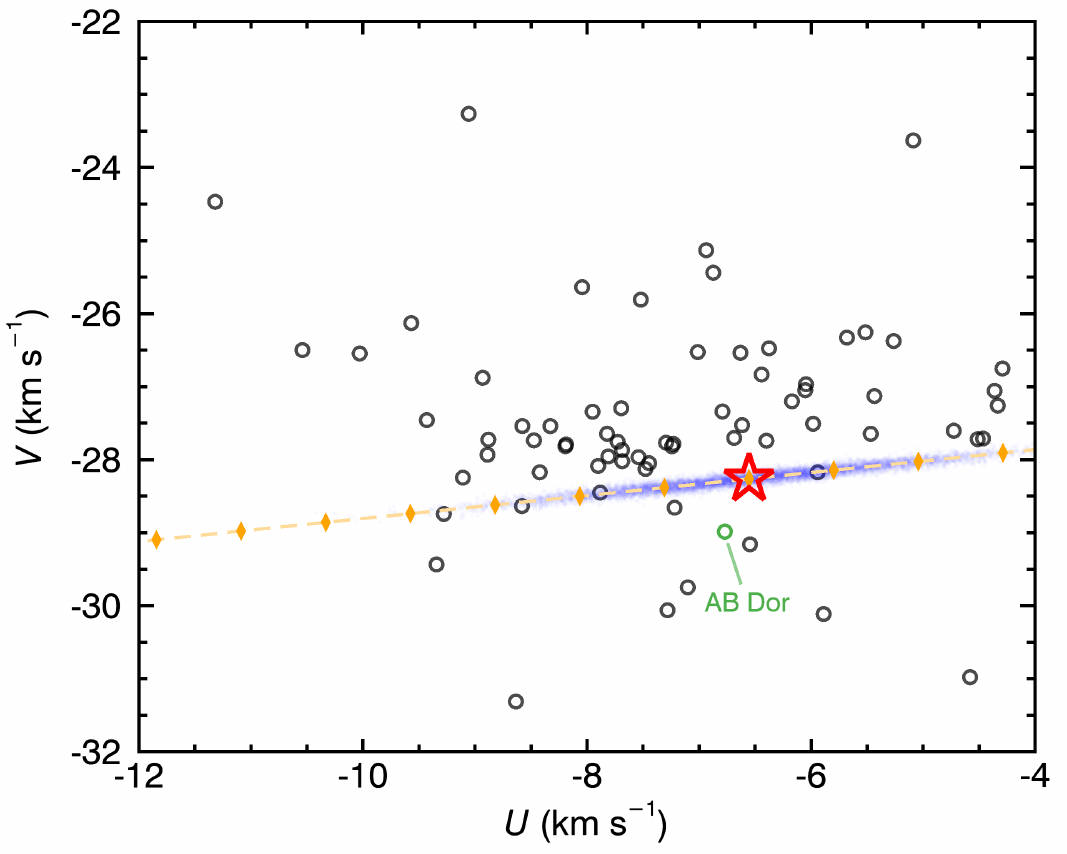}\label{fig:UV}}
	\subfigure[$VW$]{\includegraphics[width=0.48\textwidth]{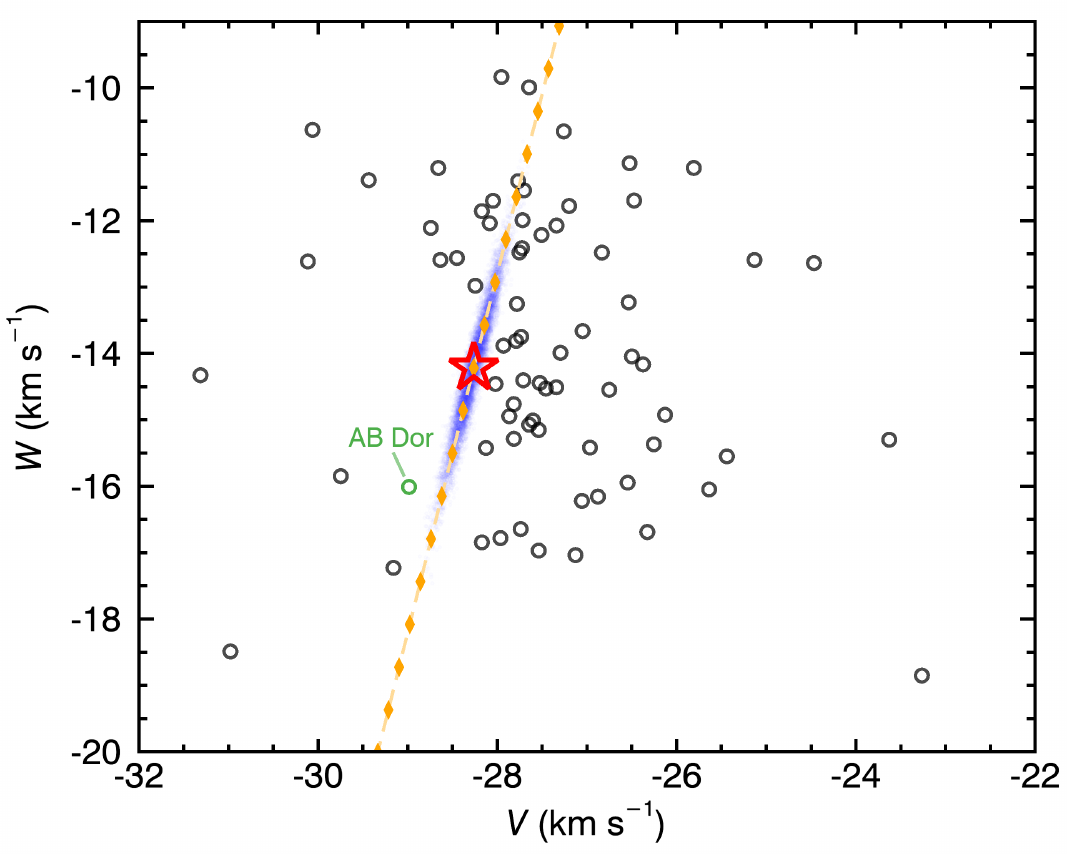}\label{fig:VW}}
	\caption{\replaced{Space velocities $UVW$}{$UVW$ space velocity} of GD~50 (red star) at its BANYAN~$\Sigma$-predicted optimal radial velocity for the ABDMG compared to the bona fide members of the ABDMG (black empty circles). The $UVW$ values of GD~50 for a range of radial velocities are displayed with the orange dashed curve, with increment of 1\,\kms\ marked as orange diamonds. The blue shaded region shows $10^4$ Monte Carlo realizations of the $UVW$ position of GD~50 adopting Gaussian error bars on its position, proper motion, parallax and optimal radial velocity. GD~50 is predicted to be located at 0.9\,\kms\ from the center of the ABDMG members locus in $UVW$ space. \added{The AB~Dor star is indicated in green, and the average $UVW$ velocity of the ABDMG members is $\left(-7.2, -27.6, -14.2\right)$\,\kms\ \citep{2018ApJ...856...23G}. }See Section~\ref{sec:ident} for more details.}
	\label{fig:kinematics}
\end{figure*}

GD~50 was recovered as a high-likelihood candidate member of the ABDMG in a search for members of young associations with the 100\,pc \emph{Gaia}~DR2 sample \citep{2018arXiv180511715G}, although this study originally excluded white dwarf candidates. This search uses the BANYAN~$\Sigma$ Bayesian membership classification algorithm \citep{2018ApJ...856...23G} to determine the probability that a given star is a member in one of the nearest 27 young associations or in the field\footnote{An IDL version is available at \url{https://github.com/jgagneastro/banyan_sigma_idl} and a Python version at \url{https://github.com/jgagneastro/banyan_sigma} \citep{zenodobanyansigmapython,zenodobanyansigmaidl}, and a web portal is available for single-object queries at \url{www.exoplanetes.umontreal.ca/banyan/banyansigma.php}}\added{, defined as all associations younger than 1\,Gyr within 150\,pc of the Sun}.

In summary, each young association is modelled with a multivariate Gaussian density in 6-dimensional Galactic position and space velocity $XYZUVW$, and the Galactic field within 300\,pc is modelled with a mixture of 10 multivariate Gaussians. BANYAN~$\Sigma$ uses Bayes' theorem to compare observables with the multivariate Gaussian models, and marginalizes over radial velocity or distance \added{-- with analytical solutions to the marginalization integrals -- }when these measurements are not available\deleted{ with an analytical solution to the marginalization integrals}. The 27 associations considered in BANYAN~$\Sigma$ include the ABDMG and the Pleiades association among others.

The \replaced{updated}{\emph{Gaia}~DR2} proper motion and parallax of GD~50 are reported in Table~\ref{tab:properties}. Analyzing these \deleted{updated }kinematics with the BANYAN~$\Sigma$ Bayesian membership classification algorithm yields a 99.7\% membership in the ABDMG, and a negligible Pleiades membership probability. BANYAN~$\Sigma$ predicts an optimal radial velocity of $17.5 \pm 1.4$\,\kms, assuming membership in the ABDMG. The resulting $UVW$ space \replaced{velocities for}{velocity of} GD~50 \replaced{are}{is} listed in Table~\ref{tab:properties}, and compared with the known members of the ABDMG in Figure~\ref{fig:kinematics}.

Measuring the radial velocity of GD~50 with a precision of a few \kms\ would be required to fully confirm its kinematic match to the ABDMG, however this is challenging because of the strong gravitational redshift at the surface of GD~50. \cite{2006MNRAS.373L..45D} measured a radial velocity of $176 \pm 4.3$\,\kms\ from the line core positions of H$\alpha$ and H$\beta$, but the correction due to gravitational \replaced{reddening}{redshift} ($162 \pm 11$\,\kms, calculated from fundamental properties given in Table~\ref{tab:properties}) makes the resulting heliocentric radial velocity ($14 \pm 12$\,\kms) highly imprecise, although consistent with the prediction of BANYAN~$\Sigma$.

\begin{figure*}
	\centering
	\subfigure[AB~Doradus moving group]{\includegraphics[width=0.48\textwidth]{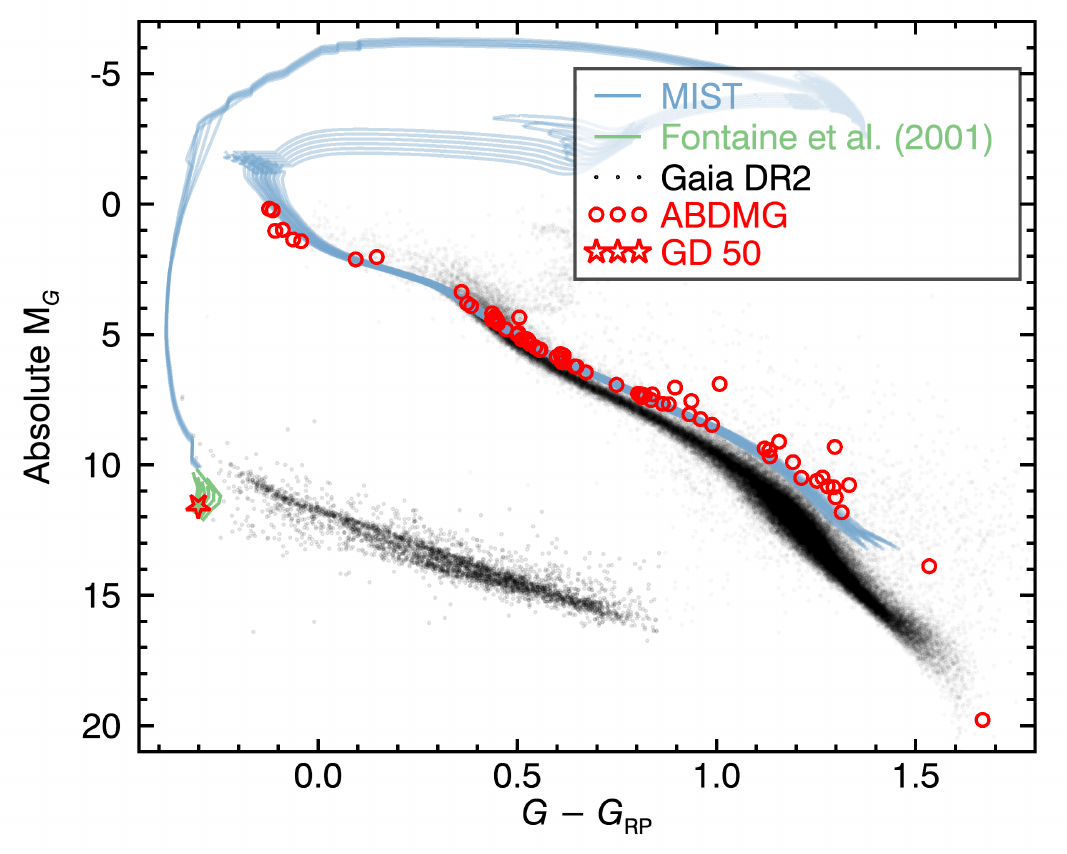}}
	\subfigure[GD~50]{\includegraphics[width=0.48\textwidth]{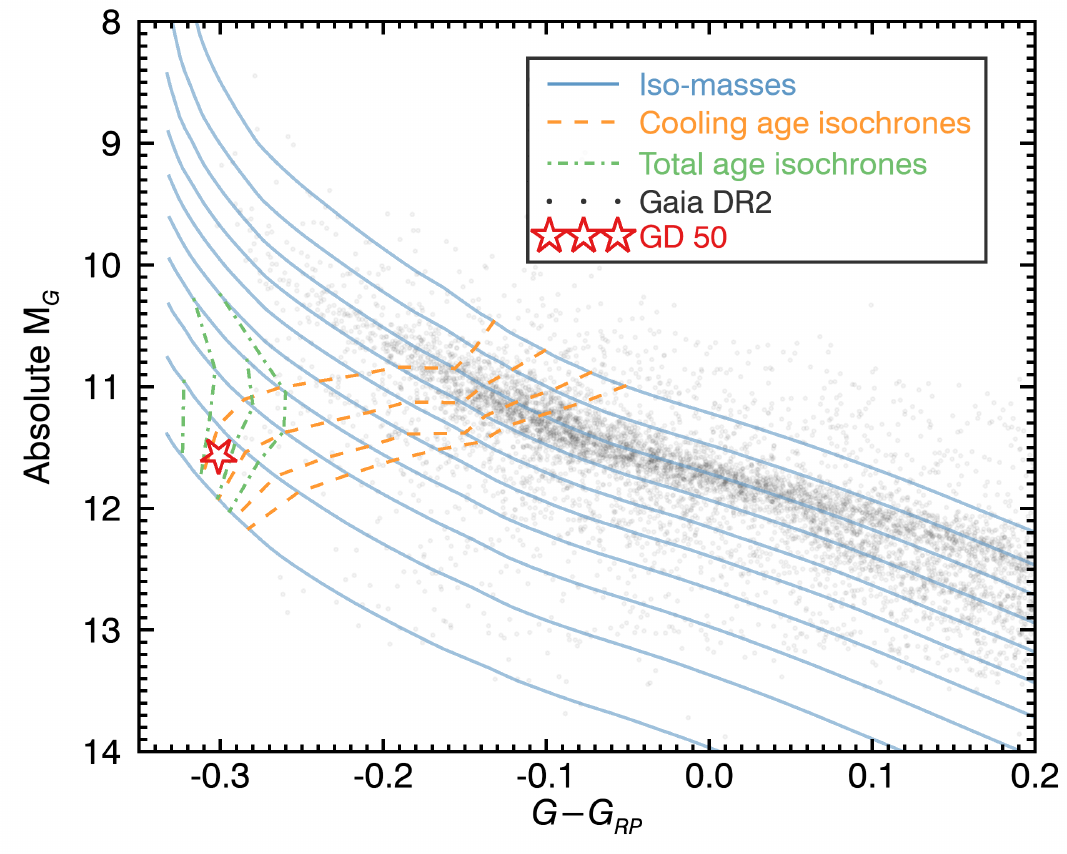}}
	\caption{Left: \emph{Gaia}~DR2 color-magnitude diagram of \replaced{the nearest 100\,pc stars}{stars within 100\,pc} (black dots) and ABDMG members (red circles and red star), compared to empirically corrected solar-metallicity MIST isochrones and the Montr\'eal white dwarf total age isochrones at 110--200\,Myr. Right: Color-magnitude diagram of the Montr\'eal C/O core white dwarf cooling tracks compared to GD~50 (red star). Iso-masses in the range 0.4--1.3\,\msol\ by steps of 0.1\,\msol are displayed as blue lines (massive white dwarfs are fainter), cooling time isochrones for ages 70, 110, 150 and 180\,Myr (left to right) are displayed with orange dashed lines, and total age isochrones for the same values are displayed with green dashed-dotted lines. See Section~\ref{sec:ident} for more details.}
	\label{fig:cmd}
\end{figure*}

In Figure~\ref{fig:cmd}, we show the \cite{2018ApJ...856...23G} compilation of bona fide members of the ABDMG in a \emph{Gaia}~DR2 absolute $G$ versus $G-G_{RP}$ color-magnitude diagram compared with the nearest 100\,pc \emph{Gaia}~DR2, the empirically corrected 110--200\,Myr MIST tracks, as well as the total white dwarf ages. The total white dwarf ages are based on MIST for the pre-white dwarf phases and the Montr\'eal C/O core cooling tracks\footnote{Computed similarly as the pure C core models described in \cite{2001PASP..113..409F}, and available at \url{http://www.astro.umontreal.ca/~bergeron/CoolingModels/}} (see also \citealp{2006AJ....132.1221H,2006ApJ...651L.137K,2011ApJ...730..128T,2011ApJ...737...28B}). GD~50 falls at a position consistent with the white dwarf isochrone tracks at the total age of the ABDMG, and forms an extension of its sequence across the stages of stellar evolution.

\section{THE INITIAL MASS FUNCTION OF THE AB~DORADUS MOVING GROUP}\label{sec:imf}

In this section we determine whether the discovery of a white dwarf member in the ABDMG is consistent with its present-day mass function. We determined the masses of bona fide members compiled by \citep{2018ApJ...856...23G} with the fiducial solar-metallicity MIST isochrones \citep{2016ApJ...823..102C} with stellar rotation ($v = 0.4$\,$v_{\rm crit}$) at 150\,Myr, consistent with the age and metallicity of the ABDMG \citep{2013ApJ...766....6B,2015MNRAS.454..593B}\footnote{\added{MIST isochrones in the \emph{Gaia} passbands are provided at \url{http://waps.cfa.harvard.edu/MIST/data/tarballs_v1.1/MIST_v1.1_vvcrit0.4_UBVRIplus.tar.gz}, based on the zero points of \cite{Evans:2018cj}}}. The individual masses were determined by selecting the nearest point on the MIST isochrone in $N\sigma$ separation. The $N\sigma$ distance\footnote{\added{We defined the $N\sigma$ distance as the quadrature sum of (1) the difference in color between GD~50 and a point on the isochrone divided by the measurement error, and (2) the similar difference in absolute magnitude between GD~50 and a point on the isochrone divided by the measurement error.}} is given by the absolute magnitude and color separations between the star and isochrone divided by the measurement errors on the position of the star in the color-magnitude diagram. We compiled these measured masses in logarithm space with a relatively large 0.4\,dex bin size to obtain a preliminary present-day mass function for the members of the ABDMG bright enough to have a parallax in \emph{Gaia}~DR2. The bin size was selected to minimize the effects of small number statistics and systematics in the model-dependent mass determinations.

The white dwarf mass of GD~50 was determined with the Montr\'eal white dwarf cooling tracks and the updated $T_{\rm eff}$ and $\log g$ reported by \cite{2011ApJ...743..138G} for GD~50 (listed in Table~\ref{tab:properties} with its other fundamental parameters). Models at various C/O/Ne core compositions that span 50\%/50\%/0\%, to 0\%/100\%/0\%, and to 0\%/0\%/100\%, were compared with the spectroscopic properties in a $10^4$-elements Monte Carlo simulation to account for measurement errors and the unknown core composition. We consider neon cores in addition to C/O cores because stellar evolution theory predicts that white dwarfs as massive as GD~50 are the progenitors of $>$\,7\,\msol\ stars, which burn oxygen in their late stages of life, and are expected to produce white dwarfs with O/Ne core compositions \citep{2013ApJ...772..150J,2015ApJ...810...34W}. We recovered a mass of $1.28 \pm 0.02$\,\msol\ with C/O core compositions, and a slightly lower mass of $1.27 \pm 0.02$\,\msol\ for a pure Ne core. Marginalizing over all possible core compositions with a uniform prior yields a mass of $1.28 \pm 0.02$\,\msol, which we adopt here. We used thick hydrogen atmosphere models ($10^{-4}$ mass fraction), but we found that using thin hydrogen atmospheres ($10^{-10}$ mass fraction) had no effect on the estimated mass.

\begin{figure}
	\centering
	\includegraphics[width=0.48\textwidth]{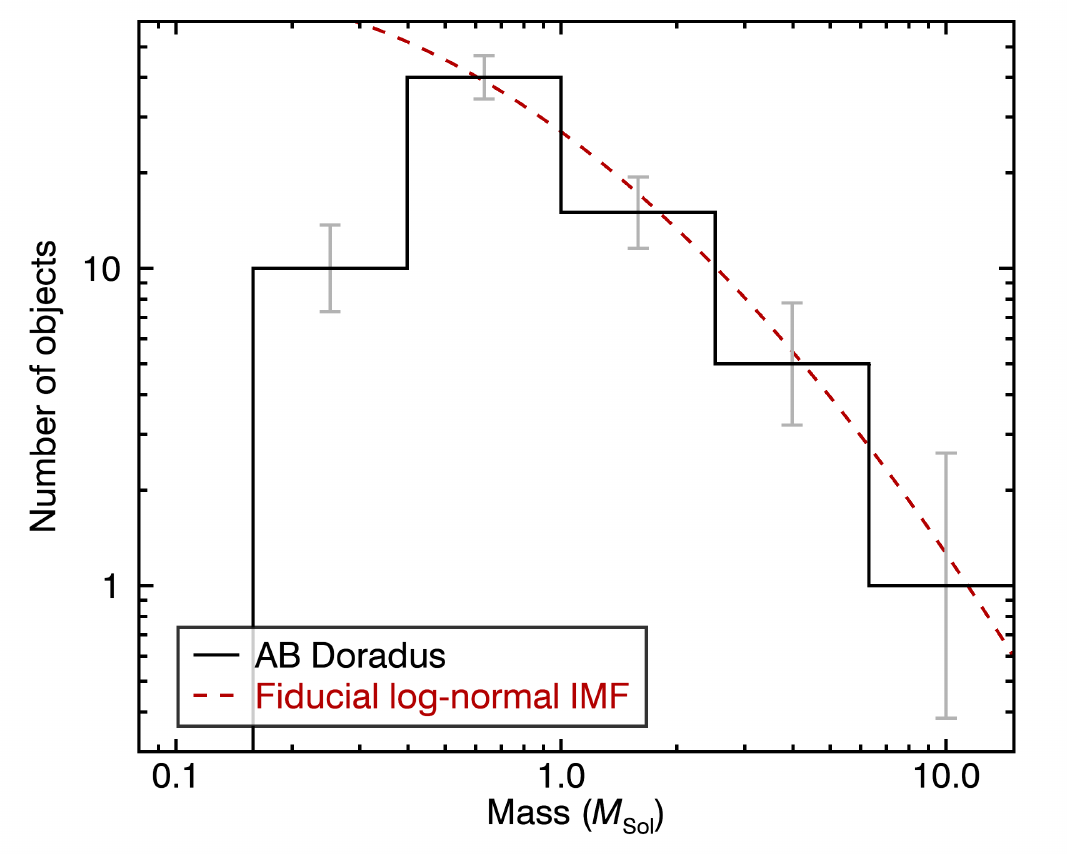}
	\caption{A preliminary initial mass function for the ABDMG that includes the $\sim$\,8\,\msol\ GD~50 progenitor (black lines). The red dashed line represents a fiducial log-normal initial mass function anchored on the AFG-type members, and the Poisson error bars on individual bins are based on small number statistics. The discovery of a single white dwarf member in the ABDMG is consistent with a log-normal initial mass function. This preliminary initial mass function is not complete in its $\sim$\,0.3\,\msol\ bin principally because of the currently incomplete census of low-mass ABDMG members with full kinematics. See Section~\ref{sec:imf} for more details.}
	\label{fig:imf}
\end{figure}

The progenitor mass of GD~50 was determined from the white dwarf mass and the initial-to-final mass relations of \cite{2013MmSAI..84...58K}\footnote{Based on the \cite{2008ApJ...676..594K} relations}. Measurement errors on the initial-to-final mass relation coefficients and the final mass were assumed to follow a Gaussian distribution and were propagated with a $10^4$ Monte Carlo analysis, resulting in a progenitor mass of \progenitormass\,\msol\ for GD~50. This mass corresponds to a main-sequence effective temperature of $\sim$\,22\,000\,K \citep{2016ApJ...823..102C}, and to a main-sequence spectral type of $\sim$\,B2 \citep{2013ApJS..208....9P}. The GD~50 progenitor mass was added to our ABDMG present-day mass function to obtain its initial mass function, displayed in Figure~\ref{fig:imf}.

A fiducial log-normal initial mass function (with a central mass $m_c = 0.1$\,\msol\ and spread $\sigma = 0.7$\,dex; \citealt{2011AJ....141...98B}) was adjusted to match the 0.6--4\,\msol\ members (corresponding to the AFG spectral classes), and is also displayed in Figure~\ref{fig:imf}. This log-normal initial mass function is consistent with $1.7_{-0.9}^{+1.9}$ white dwarf members in the ABDMG, which is indicative that we could have expected to discover a single white dwarf member at a 93\% probability, and that there is only a $\sim$\,50\% probability that at least one additional white dwarf member remains to be discovered in the region of $XYZ$ Galactic positions where the current ABDMG census of members is concentrated.

\section{THE AGE OF GD~50}\label{sec:age}

GD~50 provides a unique opportunity to use a white dwarf as a cosmochronometer to determine the age of a nearby young association. In this section, we use the Montr\'eal white dwarf cooling tracks combined with the MIST isochrones to determine the total age of GD~50, accounting for all of its stages of stellar evolution.

We used the $T_{\rm eff}$ and $\log g$ values of GD~50 listed in Table~\ref{tab:properties} to determine its cooling age. \cite{2015ASPC..493..137S} and \cite{ameliephd} demonstrated that various assumptions on the C/O core compositions can have a non-negligible effect on the cooling age of a white dwarf, depending on its mass and age. This effect could also be expected for different Ne compositions. We therefore used the Montr\'eal white dwarf cooling tracks with thick hydrogen atmospheres ($10^{-4}$ mass fraction) and various C/O/Ne core compositions that span 100\% carbon, to 100\% oxygen, and to 100\% neon to determine the cooling age of GD~50. We used a $10^4$-elements Monte Carlo simulation to propagate the measurement errors of $T_{\rm eff}$ and $\log g$. We found cooling ages of $81 \pm 13$\,Myr, $74 \pm 12$\,Myr, $69_{-11}^{+12}$\,Myr, and $65_{-9}^{+20}$\,Myr for respective core compositions of 50\%/50\% C/O, pure O, 50\%/50\% O/Ne and pure Ne. We repeated this age determination with thin hydrogen atmospheres ($10^{-10}$ mass fraction), and found that it had only a small effect, with cooling ages $\sim$\,$5$\,Myr larger than the thick hydrogen atmospheres

For each Monte Carlo realization, a progenitor mass was calculated using the \cite{2013MmSAI..84...58K} relation, and the MIST isochrones were used to determine the total time between the birth and white dwarf phase from the mass of each synthetic object. The uncertainties on the parameters of the \cite{2013MmSAI..84...58K} relation were included in the Monte Carlo analysis, and yielded typical MIST isochrone lifetimes of \msage\,Myr when marginalizing over all possible core compositions. These lifetimes were combined to the cooling ages to determine a total age of \agenosys\,Myr for GD~50 -- the core compositions that include heavier elements tend to have slightly smaller masses, and slightly longer stellar lifetimes, which counter-balances part of the uncertainty in total age caused by the cooling age--core composition correlation. Our age estimate is on the young side of recent age estimates for the ABDMG (e.g., $149_{-19}^{+51}$; \citealt{2015MNRAS.454..593B}), and consistent with the age of the Pleiades ($112 \pm 5$\,Myr; \citealt{2015ApJ...813..108D}). This is the first age determination of a nearby young moving group based on white dwarf evolution theory.

In principle, it would be possible to constrain the core composition of GD~50 by adopting an ABDMG age from an independent method (e.g., see \citealt{2015MNRAS.454..593B}), or by comparing the spectroscopic distance of different core compositions with the \emph{Gaia}~DR2 parallax measurement (using white dwarf mass-radius relations, e.g. see \citealt{2017ApJ...848...11B}). We attempted to do so, but found no useful constraints on the core composition of GD~50. For the first method this is due to the large measurement errors on the isochronal age of the ABDMG, and for the second method it is due to the error bars on $T_{\rm eff}$ and $\log g$ and the weak mass dependence on white dwarfs of C/O/Ne cores at young ages.

\added{A comparison between GD~50 and the Pleiades white dwarf LB~1497 \citep{Luyten1960} is warranted given their membership in coeval associations. LB~1497 has a lower mass of $1.05 \pm 0.03$\,\msol, with correspondingly lower effective temperature ($32700 \pm 500$\,K) and surface gravity ($8.67 \pm 0.05$\,dex; \citealt{2011ApJ...743..138G}). Using the same models and method as described above, we estimate a progenitor mass of $5.7 \pm 0.5$\,\msol\ for LB~1497, and a total age of $139_{-15}^{+18}$\,Myr. This total age is the sum of the pre-white dwarf lifetime ($88_{-16}^{+21}$\,Myr) and the cooling age ($50_{-8}^{+9}$\,Myr). This total age is slightly older than the age of the Pleiades based on the lithium depletion boundary ($112 \pm 5$\,Myr; \citealt{2015ApJ...813..108D}), which we suspect may be attributed to systematic uncertainties in the pre-white dwarf phase lifetimes that we estimated based on the MIST isochrones. We therefore assign a systematic uncertainty of $\pm 22$\,Myr to our method, to bring the disagreement between the Pleaides age and the age of LB~1497 from 1.7$\sigma$ to 1.0$\sigma$. As a consequence of this estimated systematic uncertainty, we update the age of GD~50 to \agefull\,Myr.}

Because our age measurement is based on a method independent of the \cite{2015MNRAS.454..593B} method, we can combine the two ages statistically to obtain a more precise age estimate for the ABDMG. To do so, we approximated the \cite{2015MNRAS.454..593B} measurement with an asymmetrical Gaussian probability density function and multiplied it with our age probability density function. The resulting combined age\added{ for the ABDMG} is \combinedage\,Myr.

\begin{deluxetable*}{lcc}[p]
\renewcommand\arraystretch{0.9}
\tabletypesize{\small}
\tablecaption{Properties of GD~50 \label{tab:properties}}
\tablehead{\colhead{Property} & \colhead{Value} & \colhead{Reference}}
\vspace{-0.2cm}\startdata
\sidehead{\textbf{Position and Kinematics}\vspace{-0.1cm}}
\emph{Gaia}~DR2 Source ID & 3251244858154433536 & 1\\
R.A. ep. 2015\tablenotemark{a} & 03:48:50.27 $\pm 0.05$\,mas & 1\\
Decl. ep. 2015\tablenotemark{a}  & --00:58:34.8 $\pm 0.04$\,mas & 1\\
R.A. ep. 2000\tablenotemark{b} & 03:48:50.19 $\pm 1.6$\,mas & 1\\
Decl. ep. 2000\tablenotemark{b}  & --00:58:32.4 $\pm 1.3$\,mas & 1\\
$\mu_\alpha\cos\delta$ (\masyr) & $84.4 \pm 0.1$ & 1\\
$\mu_\delta$ (\masyr) & $-162.96 \pm 0.09$ & 1\\
Parallax (mas) & $32.04 \pm 0.06$ & 1\\
Trigonometric distance (pc) & $31.21 \pm 0.06$ & 1\\
RV$_{\rm opt}$\tablenotemark{c} (\kms) & $17.5 \pm 1.4$ & 2\\
RV$_{\rm mes}$\tablenotemark{c} (\kms) & $14 \pm 12$ & 3\\
$X$ (pc) & $-23.59 \pm 0.04$ & 2\\
$Y$ (pc) & $-3.717 \pm 0.007$ & 2\\ 
$Z$ (pc) & $-20.10 \pm 0.04$ & 2\\
$U$\tablenotemark{d} (\kms) & $-7 \pm 1$ & 2\\
$V$\tablenotemark{d} (\kms) & $-28.3 \pm 0.2$ & 2\\
$W$\tablenotemark{d} (\kms) & $-14.2 \pm 0.9$ & 2\vspace{-0.2cm}\\
\sidehead{\textbf{Photometric Properties}\vspace{-0.1cm}}
$B$ (Johnson) & $14.063 \pm 0.0032$ & 4\\
$B-V$ (Johnson) & $-0.276 \pm 0.0036$ & 4\\
$G_{\rm BP}$ (\emph{Gaia}~DR2) & $13.782 \pm 0.007$ & 1\\
$G$ (\emph{Gaia}~DR2) & $14.0354 \pm 0.0005$ & 1\\
$G_{\rm RP}$ (\emph{Gaia}~DR2) & $14.336 \pm 0.001$ & 1\\
$u_{\rm AB}$ (SDSS~DR12) & $13.396 \pm 0.003$ & 5\\
$g_{\rm AB}$ (SDSS~DR12) & $13.760 \pm 0.003$ & 5\\
$r_{\rm AB}$ (SDSS~DR12) & $14.282 \pm 0.004$ & 5\\
$i_{\rm AB}$ (SDSS~DR12) & $14.640 \pm 0.004$ & 5\\
$z_{\rm AB}$ (SDSS~DR12) & $14.972 \pm 0.005$ & 5\\
$g_{\rm AB}$ (Pan-STARRS~DR1) & $13.831 \pm 0.004$ & 6\\
$r_{\rm AB}$ (Pan-STARRS~DR1) & $14.290 \pm 0.002$ & 6\\
$i_{\rm AB}$ (Pan-STARRS~DR1) & $14.678 \pm 0.003$ & 6\\
$z_{\rm AB}$ (Pan-STARRS~DR1) & $14.972 \pm 0.003$ & 6\\
$y_{\rm AB}$ (Pan-STARRS~DR1) & $15.168 \pm 0.004$ & 6\\
$J$ (2MASS) & $14.75 \pm 0.03$ & 7\\
$H$ (2MASS) & $14.86 \pm 0.04$ & 7\\
$K_S$ (2MASS) & $15.120 \pm 0.138$ & 7\\
$J$ (UKIDSS~DR9) & $14.794 \pm 0.004$ & 8\\
$H$ (UKIDSS~DR9) & $14.890 \pm 0.007$ & 8\\
$K$ (UKIDSS~DR9) & $15.05 \pm 0.01$ & 8\\
$W1$ (AllWISE) & $15.11 \pm 0.04$ & 9\\
$W2$ (AllWISE) & $15.20 \pm 0.09$ & 9\\
\sidehead{\textbf{Fundamental Properties}\vspace{-0.1cm}}
Spectral type & DA1.2 & 10\\
\teff\ (K) & $42700 \pm 800$ & 10\\
$\log g$ & $9.20 \pm 0.07$ & 10\\
Mass (\msol) & $1.28 \pm 0.02$ & 2\\
Age (Myr) & \agefull & 2\\
\enddata
\tablenotetext{a}{J2000 position at epoch 2015 from the \emph{Gaia}~DR2 catalog.}
\tablenotetext{b}{J2000 position at epoch 2000 calculated from the \emph{Gaia}~DR2 astrometric solution.}
\tablenotetext{c}{Optimal radial velocity predicted by BANYAN~$\Sigma$ that assumes the most likely membership hypothesis (ABDMG).}
\tablenotetext{d}{$UVW$ values presented here assume the optimal radial velocity produced by BANYAN~$\Sigma$ that maximizes the ABDMG membership probability.}
\tablerefs{(1)~\citealt{Lindegren:2018gy}, (2)~This work, (3)~\citealt{2006MNRAS.373L..45D}, (4)~\citealt{2009AJ....137.4186L}, (5)~\citealt{2015ApJS..219...12A}, (6)~\citealt{2016arXiv161205560C}, (7)~\citealt{2006AJ....131.1163S}, (8)~\citealt{2007MNRAS.379.1599L}, (9)~\citealt{2014ApJ...783..122K}, (10)~\citealt{2011ApJ...743..138G}.
}
\end{deluxetable*}

\section{CONCLUSION}\label{sec:conclusion}

We present evidence that the ultramassive white dwarf GD~50 is a member of the ABDMG based on its updated \emph{Gaia}~DR2 kinematics and the BANYAN~$\Sigma$ bayesian classification algorithm. \added{We estimate a mass of \progenitormass\,\msol\ for the progenitor of GD~50, and find that }\replaced{A}{a} log-normal initial mass function anchored on the AFG-type members of the ABDMG is consistent with at least one white dwarf member at a 93\% statistical confidence. The inclusion of GD~50 in the list of ABDMG members makes it possible to use white dwarf cooling ages as a new method to constrain the age of ABDMG, independent of the lithium depletion boundary and isochrone methods. We use this method to find an age of \age\,Myr, which is on the younger side of recent age determinations based on isochrones. Neither the \emph{Gaia}~DR2 parallax nor literature ages for the ABDMG can constrain the relative compositions of C/O/Ne in the core of GD~50.

Our analysis corroborates the conclusion of \cite{2006MNRAS.373L..45D} that a single massive progenitor can form white dwarfs as massive as GD~50 without the need of invoking white dwarf mergers which simulations have difficulty producing at a high enough rate to explain the number of observed ultramassive white dwarfs \citep{1997ApJ...481..355S}. Identifying additional white dwarfs in young associations of the Solar neighborhood will provide powerful constraints on their ages which are independent of other widely used methods.

\acknowledgments

\added{We thank the anonymous reviewer for insightful comments that improved the quality of this manuscript. }We thank Ricky L. Smart for useful comments and Pierre Bergeron for sending white dwarf model tracks in the \emph{Gaia}~DR2 passbands. This research made use of: the SIMBAD database and VizieR catalog access tool, operated at the Centre de Donn\'ees astronomiques de Strasbourg, France; data products from the Two Micron All Sky Survey, which is a joint project of the University of Massachusetts and the Infrared Processing and Analysis Center (IPAC)/California Institute of Technology (Caltech), funded by the National Aeronautics and Space Administration (NASA) and the National Science Foundation; data products from the \emph{Wide-field Infrared Survey Explorer}, which is a joint project of the University of California, Los Angeles, and the Jet Propulsion Laboratory at Caltech, funded by NASA. This work presents results from the European Space Agency (ESA) space mission \emph{Gaia}. \emph{Gaia} data are being processed by the \emph{Gaia} Data Processing and Analysis Consortium (DPAC). Funding for the DPAC is provided by national institutions, in particular the institutions participating in the \emph{Gaia} MultiLateral Agreement (MLA). The \emph{Gaia} mission website is https://www.cosmos.esa.int/gaia. The \emph{Gaia} archive website is https://archives.esac.esa.int/gaia. This research was started at the NYC \emph{Gaia} DR2 Workshop at the Center for Computational Astrophysics of the Flatiron Institute in 2018 April.

\software{BANYAN~$\Sigma$ \citep{2018ApJ...856...23G}.}


\end{document}